\documentclass{article}
\usepackage{amssymb}
\usepackage{amstext}
\usepackage{color}
\usepackage{graphicx}
\usepackage{url}
\usepackage{float}
\usepackage{sidecap}
\usepackage{subfig}
\usepackage{listings}
\newcommand{\likwid}{LIKWID}

\newcommand{\bq}{\begin{equation}}
\newcommand{\eq}{\end{equation}}

\newcommand{\flop}{\mbox{flop}}
\newcommand{\flops}{\mbox{flops}}

\newcommand{\GBS}{\mbox{GB/s}}

\newcommand{\GFS}{\mbox{GFlop/s}}

\newcommand{\GHZ}{\mbox{GHz}}

\newcommand{\bytes}{\mbox{bytes}}

\newcommand{\MB}{\mbox{MB}}

\newcommand{\eos}{~.}

\begin{document}
\title{\bfseries Hybrid-parallel sparse matrix-vector multiplication
with explicit communication overlap on current multicore-based systems}
\author{Gerald Schubert, Holger Fehske\\
Institute of Physics, University of Greifswald\\
Felix-Hausdorff-Str. 6, 17487 Greifswald, Germany\\[4mm]
Georg Hager, Gerhard Wellein\\
Erlangen Regional Computing Center, University of Erlangen-Nuremberg\\
Martensstr. 1, 91058 Erlangen, Germany}
\date{June 14, 2011}
\maketitle
\begin{abstract}
  We evaluate optimized parallel sparse matrix-vector operations for
  several representative application areas on widespread multicore-based
  cluster configurations.  First the single-socket baseline
  performance is analyzed and modeled with respect to basic
  architectural properties of standard multicore chips. Beyond
  the single node, the performance of parallel sparse matrix-vector operations 
  is often limited by communication overhead. Starting from the observation 
  that nonblocking MPI is not
  able to hide communication cost using standard MPI implementations,
  we demonstrate that explicit overlap of communication and
  computation can be achieved by using a dedicated communication
  thread, which may run on a virtual core. Moreover we identify
  performance benefits of hybrid MPI/OpenMP programming
  due to improved load balancing even without explicit communication
  overlap. 
  We compare performance results for pure MPI, the
  widely used ``vector-like'' hybrid programming strategies,
  and explicit overlap on a modern multicore-based cluster 
  and a Cray XE6 system.
\end{abstract}

\section{Introduction}

Many problems in science and engineering involve the solution of large
eigenvalue problems or extremely sparse systems of linear equations
arising from, e.g., the discretization of partial differential
equations. Sparse matrix-vector multiplication (spMVM) is the dominant
operation in many of those solvers and may easily consume most of the
total run time.
A highly efficient scalable spMVM implementation is thus fundamental,
and complements advancements and new development in the high-level
algorithms.

For more than a decade there
has been an intense debate about whether the hierarchical structure
of current HPC systems needs to be considered in parallel programming, 
or if pure MPI is
sufficient. Hybrid approaches based on MPI+OpenMP have
been implemented in codes and kernels for various applications areas
and compared with traditional MPI implementations. Most results are
hardware-specific, and sometimes contradictory.
In this paper we analyze hybrid MPI+OpenMP variants
of a general parallel spMVM operation. Beyond the naive
approach of using OpenMP for parallelization of kernel
loops (``vector mode'') we also employ a hybrid
``task mode''  to overcome or mitigate a weakness of
standard MPI implementations: the lack of
truly asynchronous communication in nonblocking
MPI calls. 
We test our
implementation against pure MPI approaches for two application
scenarios on an InfiniBand cluster as well as a Cray XE6 system.

\subsection{Related work}

In recent years the performance of various spMVM algorithms has been
evaluated by several groups~\cite{h0441_GKAKK08,h0441_WOVSYD09,h0441_BG09p}.
Covering different matrix storage formats and implementations on
various types of hardware, they have reviewed a more or less large number
of publicly available matrices and reported on the obtained
performance.
Scalable parallel spMVM implementations have also been
proposed \cite{symspmvm10,spmvmgr01}, mostly
based on an MPI-only strategy. Hybrid parallel spMVM approaches have
already been devised before the emergence of multicore
processors~\cite{rw03,vecpar02}. Recently a ``vector mode'' approach could not
compete with a scalable MPI implementation for a specific
problem on a Cray system~\cite{symspmvm10}. There is no up-to-date
literature that systematically investigates novel features like
multicore, ccNUMA node structure, and simultaneous multithreading
(SMT) for hybrid parallel spMVM.

\subsection{Sparse matrix-vector multiplication and node-level performance model}
\label{sect:perfmod}

\begin{lstlisting}[float=tb,caption={CRS sparse matrix-vector multiplication kernel},label=lst:crs]
  do i = 1,%$N_\mathrm{r}$%
    do j = row_ptr(i), row_ptr(i+1) - 1
      C(i) = C(i) + val(j) * %B(col\verb._.idx(j))%
    enddo
  enddo
\end{lstlisting}
A possible definition of a  ``sparse'' matrix is that the number of 
its nonzero entries grows only linearly 
with the matrix dimension; however, not all problems are easily scaled, so in general
a sparse matrix may be defined to contain ``mainly'' zero entries. Since keeping such a
matrix in computer memory with all zeros included is usually out of the
question, an efficient format to store the nonzeros only is required.
The most widely used variant is ``Compressed Row Storage''
(CRS)~\cite{barrett:1994}.
It does not exploit specific features that
may emerge from the underlying physical problem like, e.g., 
block structures, symmetries, etc., but is broadly
recognized as the most efficient format 
for general sparse matrices on cache-based microprocessors. All nonzeros are
stored in one contiguous array \verb.val(:)., row by row, and the
starting offsets of all rows are contained in a separate array
\verb.row_ptr(:).. Array \verb.col_idx(:). contains the original column
index of each matrix entry.
A matrix-vector multiplication with a right-hand-side (RHS) vector \verb.B(:).
and a result vector \verb.C(:). can then be written as shown
in Listing~\ref{lst:crs}.
Here $N_\mathrm r$ is the number of matrix rows. While arrays \verb.C(:).
and \verb.val(:). are traversed contiguously, access to \verb.B(:). is
indexed and may potentially cause very low spatial and temporal
locality in this data stream.


The performance of spMVM operations on a single compute
node is often limited by main memory 
bandwidth. Code balance~\cite{hpc4se} is thus a good metric to establish a simple
performance model. We assume the average length of the inner ($j$) loop to
be $N_\mathrm{nzr}=N_\mathrm{nz}/N_\mathrm r$, 
where $N_\mathrm{nz}$ is the total number of
nonzero matrix entries. Then the contiguous data
accesses in the CRS code generate $(8+4+16/N_\mathrm{nzr})$ \bytes{} of 
memory traffic for a single inner loop iteration, where the first two contributions come
from the matrix \verb.val(:). (8\,\bytes) and the index array
\verb.col_idx(:). (4\,\bytes), while the last term reflects the
update of \verb.C(i). (write allocate + evict). 
The indirect access pattern to \verb.B(:). is determined by the
sparsity structure of the matrix and can not be modeled in general. 
However, \verb.B(:). needs to be loaded at least once from main
memory, which adds another $8/N_\mathrm{nzr}$\,\bytes{} per inner
iteration. Limited cache size and nondiagonal access typically require
loading at least parts of \verb.B(:). multiple times in a single MVM. 
This is quantified by a machine- and problem-specific parameter $\kappa$:
For each additional time that \verb.B(:). is loaded from main memory,
$\kappa$ increases by $8/N_\mathrm{nzr}$.
Together with the
computational intensity of 2\,\flops\ per $j$ iteration the code balance is
\bq\label{eq:pmodel}
B_\mathrm{CRS}=\left(\frac{12 + 24/N_\mathrm{nzr} + \kappa}{2}\right)\frac{\bytes}{\flop}
=\left(6+\frac{12}{N_\mathrm{nzr}}+\frac{\kappa}{2}\right) \frac{\bytes}{\flop}\eos
\eq
On the node level $B_\mathrm{CRS}$ can be used to determine an upper
performance limit by measuring the node memory bandwidth (e.g., using 
the STREAM benchmark) and assuming
$\kappa=0$. Moreover, $\kappa$ can be
determined experimentally from the spMVM floating point performance
and the memory bandwidth drawn by the CRS code (see Sect.~\ref{sec:nodeperf}). Since
the ``slimmest'' matrices used here have $N_\mathrm{nzr} \approx 7\ldots 15$, 
each additional access to
\verb.B(:). incurs a nonnegligible contribution to the data
transfer in those cases.

Note that this simple model neglects performance-limiting aspects
beyond bandwidth bottlenecks like in-cache transfer time, 
load imbalance, communication
and/or synchronization overhead, and the adverse effects of
nonlocal memory access across ccNUMA locality domains (LDs).

\subsection{Experimental setting}

\subsubsection{Test matrices}\label{sec:matrices}

\begin{figure}[tb]
\includegraphics*[width=\textwidth]{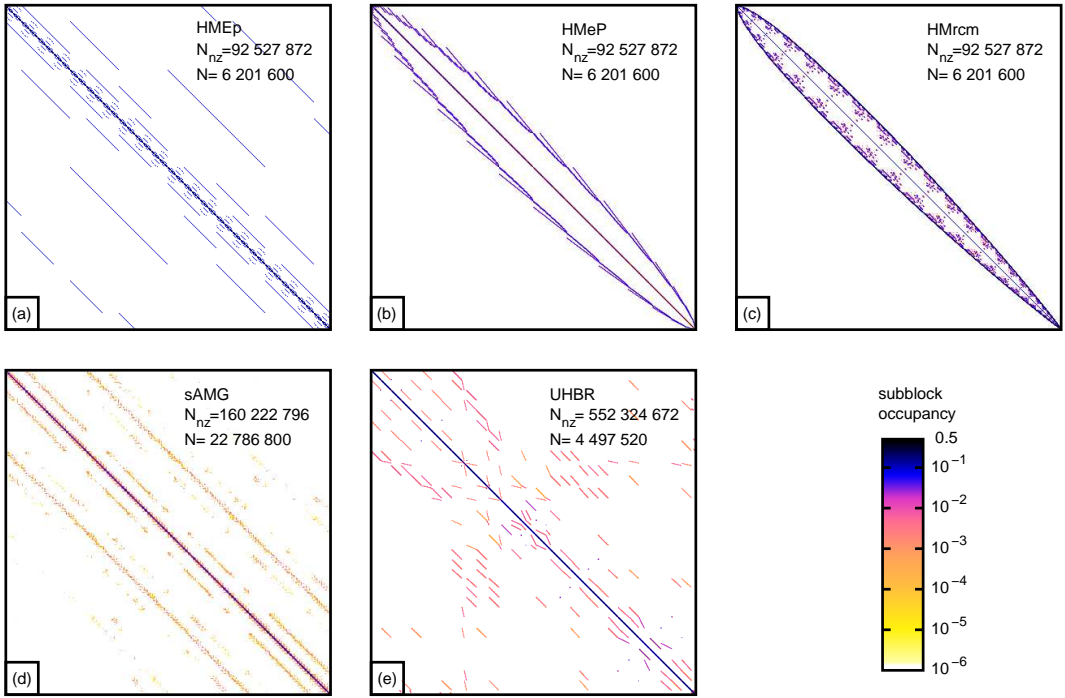}
\caption{Sparsity patterns of the matrices described in Sect.~\ref{sec:matrices}.
 (a)--(c) describe the same physical system, but use a different numbering of the
 basis elements. See text for details.
 Square subblocks
 have been aggregated and color-coded according to occupancy to improve visibility.
}\label{fig:mat}
\end{figure}
Since the
sparsity pattern may have substantial impact on the single
node performance and parallel scalability, we have chosen three
application areas known to generate extremely sparse matrices. 

As a first test case we use a matrix from exact diagonalization
of strongly correlated electron-phonon systems in solid state
physics. Here generic
microscopic models are used to treat both charge (electrons) and
lattice (phonons) degrees of freedom in second quantization. Choosing
a finite-dimensional basis set, which is the direct product of basis
sets for both subsystems (electrons $\otimes$ phonons), the generic
model can be represented by a sparse Hamiltonian matrix. 
Iterative algorithms such as Lanczos or Jacobi-Davidson are used to
compute low-lying eigenstates of the Hamilton matrices, and more recent
methods based on polynomial expansion 
allow for computation of spectral properties~\cite{WWAF06} or time
evolution of quantum states~\cite{WF06}. In all those algorithms,
spMVM is the most time-consuming step.

\begin{figure}[tb]
\subfloat[Intel dual Westmere node with two NUMA locality domains]{\includegraphics[height=0.32\linewidth]{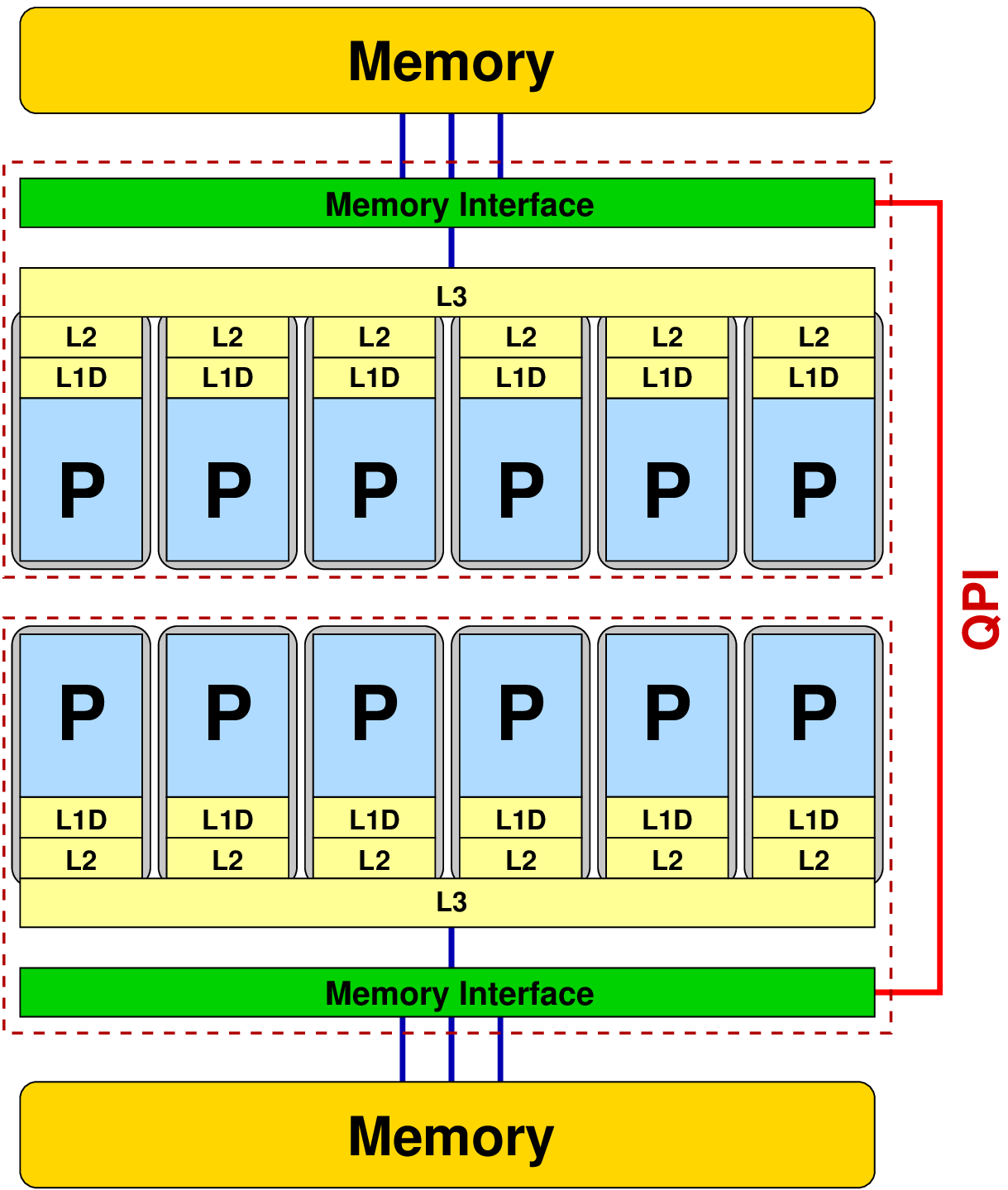}}\hfill
\subfloat[Cray XE6/AMD dual Magny Cours node with four NUMA locality domains]{\includegraphics[height=0.32\linewidth]{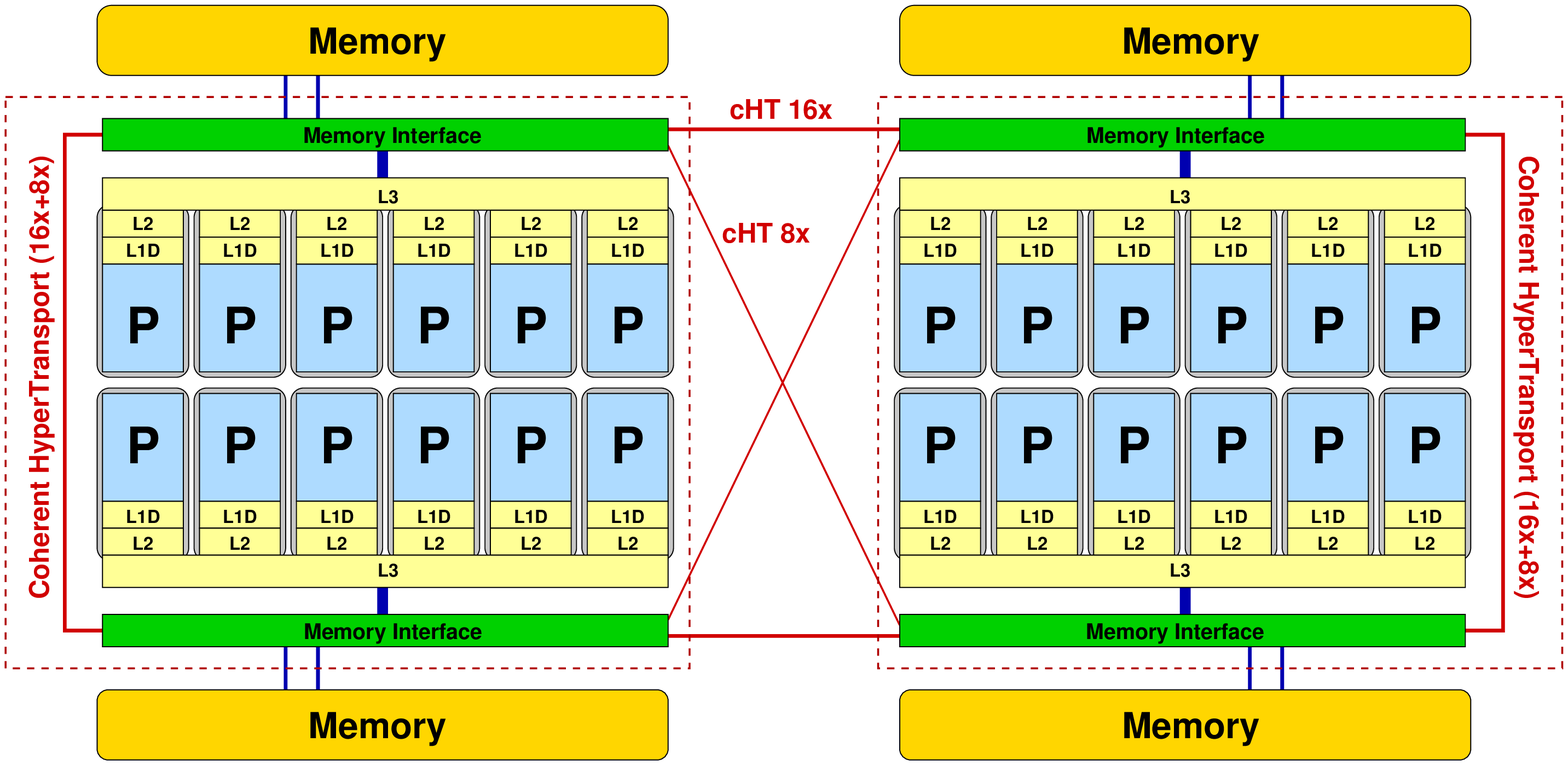}}
\caption{\label{fig:hardware}Node topology of the benchmark systems. Dashed boxes indicate sockets.}
\end{figure}
In this paper we consider the Holstein-Hubbard model
(cf.~\cite{fwhwb04} and references therein) and choose six electrons
(subspace dimension 400) on a six-site lattice coupled to 15 phonons
(subspace dimension $1.55\times10^4$). The resulting matrix of
dimension $6.2\times 10^6$ is very sparse ($N_\mathrm{nzr}\approx 15$)
and can have two different sparsity patterns, depending on whether the
phononic or the electronic basis elements are numbered contiguously
(see Figs.~\ref{fig:mat}\,(a) and (b), respectively). 
We also applied the well-known ``Reverse Cuthill-McKee (RCM)''
algorithm~\cite{RCM69} to the Hamilton matrix  in order to improve spatial 
locality  in the access to the right hand side vector,
and to optimize interprocess communication patterns towards near-neighbor
exchange.
Since the RCM-optimized
structure (Fig.~\ref{fig:mat}\,(c))  showed no performance advantage over 
the HMeP variant (Fig.~\ref{fig:mat}\,(b))
neither on the node nor on the highly parallel level,
we will not consider RCM any further in the following.

The second matrix was generated by the adaptive multigrid code sAMG
(see~\cite{AMG,sAMG}, and references therein) for the irregular discretization
of a Poisson problem on a car geometry. Its matrix dimension is
$2.2\times10^7$ with an average of $N_\mathrm{nzr}\approx 7$ entries
per row (see Fig.~\ref{fig:mat}\,(d)).

The UHBR matrix (see Fig.~\ref{fig:mat}\,(e)) originates from
aeroelastic stability investigations of an ultra-high bypass ratio
(UHBR) turbine fan of the German Aerospace Center (DLR) with a
linearized Navier-Stokes solver~\cite{UHBR}. This solver is part of
the parallel simulation system TRACE (Turbo-machinery Research
Aerodynamic Computational Environment) which was developed by DLR's
Institute for Propulsion Technology. Its matrix dimension is
$4.5\times10^6$ with an average of $N_\mathrm{nzr}\approx 123$ entries
per row, making it a rather `densely populated' sparse matrix in
comparison to the other test cases.


For symmetric matrices as considered here it would be
sufficient to store the upper triangular matrix elements and perform,
e.g., a parallel symmetric CRS spMVM~\cite{symspmvm10}. The data
transfer volume is then reduced by almost a factor of two, allowing for
a corresponding performance improvement. We do
not use this optimization here for two major reasons. First, the
discussion of the hybrid parallel vs. MPI-only implementation should
not be restricted to the special case of explicitly symmetric
matrices. Second, an efficient shared-memory
implementation of a symmetric CRS spMVM base routine has 
been presented only very recently~\cite{symmvmB}.

\subsubsection{Test machines}

\paragraph{Intel Nehalem EP / Westmere EP}

The two Intel platforms
represent a ``tick'' step within Intel's ``tick-tock'' product
strategy. Both processors only differ in a few microarchitectural
details; the most important difference is that Westmere, due
to the 32\,nm production process, accommodates
six cores per socket instead of four while keeping the same
L3 cache size per core (2\,\MB) as Nehalem. 
The processor chips (Xeon X5550 and X5650) used for the benchmarks
run at 2.66\,\GHZ{} base frequency with
``Turbo Mode'' and Simultaneous Multithreading (SMT) enabled.
A single socket forms its own ccNUMA LD via  three
DDR3-1333 memory channels (see Fig.~\ref{fig:hardware}\,(a)), allowing
for a peak bandwidth of 32\,\GBS. We use standard dual-socket nodes
that are connected via fully nonblocking QDR InfiniBand (IB)
networks. The Intel compiler in version 11.1 and the Intel MPI library
in version 4.0.1 were used throughout. Thread-core affinity was
controlled with the \likwid~\cite{likwid} toolkit.

\paragraph{Cray XE6 / AMD Magny Cours}

The Cray XE6 system is based on dual-socket nodes with AMD Magny Cours 
12-core processors (2.1\,\GHZ{} Opteron 6172) and the latest Cray ``Gemini''
interconnect. The internode bandwidth of the 2D torus network is 
beyond the capability of QDR InfiniBand.  The single node architecture depicted in
Fig.~\ref{fig:hardware}(b) reveals a unique feature of the AMD Magny
Cours chip series: The 12-core package comprises two 6-core chips
with separate L3 caches and memory controllers, tightly bound by
``1.5'' HyperTransport (HT) 16x links. Each 6-core unit forms its own NUMA
LD via two DDR3-1333 channels, i.e.,
a two-socket node comprises four NUMA locality domains.
In total the AMD design uses eight memory
channels, 
allowing for a theoretical main memory bandwidth advantage of $8/6$
over a Westmere node.  The Cray compiler in version 7.2.8 was used for
the Cray/AMD measurements. 

In Sect.~\ref{sec:nbmpi} we also show MPI performance results for
an older Cray XT4 system based on AMD Opteron ``Barcelona''
processors. 

\section{Node-level performance analysis}\label{sec:nodeperf}

\begin{figure}[tb]
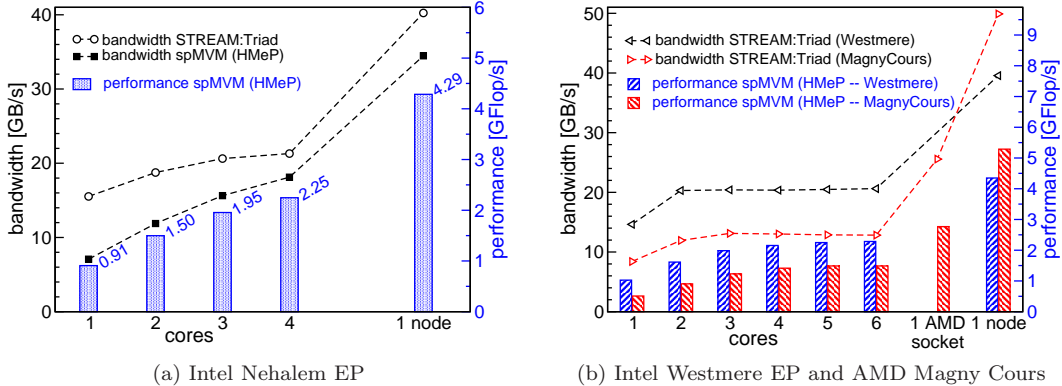

\subfloat[Intel Nehalem EP]{\includegraphics*[width=0.48\textwidth]{Neha_Node}}\hfill
\subfloat[Intel Westmere EP and AMD Magny Cours]{\includegraphics*[width=0.48\textwidth]{MaCoWest_Node}}\hfill
\caption{Node-level performance for the test systems. Effective STREAM 
triads bandwidth$^\mathrm{a}$, and performance for 
spMVM using the HMeP matrix (bars) is shown. In (a) we also report the 
measured memory bandwidth for the spMVM operation.}\label{fig:nodeperf}
\end{figure}
The basis for each parallel program must be an efficient single
core/node implementation. Assuming general sparse matrix structures, the CRS
format presented above is very suitable for modern
cache-based multicore processors~\cite{SHF09}. Even advanced 
machine-specific optimizations such as nontemporal prefetch
instructions for Opteron
processors provide only minor benefits~\cite{symspmvm10} and are thus
not considered here. A simple OpenMP parallelization of the
outermost loop, together with an appropriate NUMA-aware data
placement strategy has proven to provide best
node-level performance. We choose the HMeP, HMEp, and UHBR
matrices as reference cases for our performance model. 

Intrasocket and intranode spMVM scalability
should always be discussed together with effective STREAM triad
numbers, which form a practical upper bandwidth 
limit.\footnote{Nontemporal stores have been suppressed in 
the STREAM measurements  and the bandwidth numbers reported have been scaled
appropriately ($ \times 4/3$) to account for the write-allocate transfer.} 
Figure~\ref{fig:nodeperf}\,(a) shows the memory bandwidth on the
Nehalem EP platform drawn by the STREAM triad and the spMVM as measured 
with \likwid~\cite{likwid}.
While the STREAM bandwidth soon saturates within a socket, the spMVM
bandwidth and the corresponding \GFS\ numbers still benefit from
the use of all cores. This is a typical behavior for codes with 
(partially) irregular data access patterns. However, the fact that
more than 85\% of the STREAM bandwidth can be reached with spMVM
indicates that our CRS implementation makes good use of the
resources. The maximum spMVM performance can be estimated by dividing
the memory bandwidth by the code balance (\ref{eq:pmodel}), using $N_\mathrm{nzr}=15$
and $\kappa=0$. For a single socket the spMVM draws 18.1\,\GBS{}
(STREAM triads: 21.2\,\GBS{}), allowing for a maximum performance of
2.66\,\GFS{} (3.12\,\GFS{}). 
Combining the measured performance (2.25\,\GFS{}) and
bandwidth of the spMVM operation with $B_\mathrm{CRS}(\kappa)$ we find
$\kappa=2.5$, i.e., 2.5 additional \bytes\ of memory traffic on \verb.B(:).
per inner loop iteration  (37.3\,\bytes\ per row) are 
required due to limited cache capacity. Thus the complete vector \verb.B(:). is
loaded six times from main memory to cache, but each element is used
$N_\mathrm{nzr}=15$ times on average. This ratio
gets worse if the matrix bandwidth increases. For the HMEp matrix we found
$\kappa=3.79$, which translates to a 50\% increase in the additional
data transfers for \verb.B(:).. The code balance implies a performance
drop of about 10\%, which is consistent with our measurements.

The UHBR matrix represents an interesting case, since the average
number of nonzeros per row is $N_\mathrm{nzr}\approx 123$. At a measured
spMVM bandwidth of 18.9\,\GBS\ and a performance of 2.99\,\GFS\ per socket
we arrive at $\kappa=0.43$, which means
that each element of the RHS is loaded 8 times; however, it is used
123 times, which leads to the conclusion that the contribution of the
RHS to the memory traffic is minor. We have included this example
here because it shows that the data transfer for the RHS may be
negligible even if it is loaded many times  --- $N_\mathrm{nzr}$
plays a decisive role. Nevertheless, since this matrix shows 
perfect scaling in the highly parallel case we will not discuss
it any further in this work.

In Fig.~\ref{fig:nodeperf}\,(b) we summarize the performance
characteristics for Intel Westmere and AMD Magny Cours, which
both comprise six cores per locality domain. While the
AMD system is slower on a single LD, its node-level performance
is about 25\% higher than on Westmere due to its four LDs
per node. Within the domains spMVM saturates at four cores on
both architectures, leaving ample room to use the remaining cores for
other tasks, like communication (see Sect.~\ref{sec:taskmode}).
In the following we will report results for the Westmere and Magny Cours 
platforms only.

\section{Distributed-memory parallelization}\label{sec:dmpar}

\subsection{Nonblocking point-to-point communication in MPI}
\label{sec:nbmpi}

Strong scaling of MPI-parallel spMVM is inevitably limited
by communication overhead. Hence, it is vital to find ways to
hide communication costs as far as possible. A widely used
approach is to employ nonblocking point-to-point MPI calls
for overlapping communication with useful work. 
However, it has been known for a long time that most MPI
implementations support progress, i.e., actual data transfer, only 
when MPI library code is executed by the user process, although the
hardware even on standard InfiniBand-based clusters does
not hinder truly asynchronous point-to-point communication.\footnote{%
In fact, dedicated ``offload'' communication hardware was not unusual
in historic supercomputer architectures, the Intel Paragon of the early
1990s being a typical example.} 

\begin{SCfigure}[0.38][b]
\includegraphics*[width=0.7\linewidth]{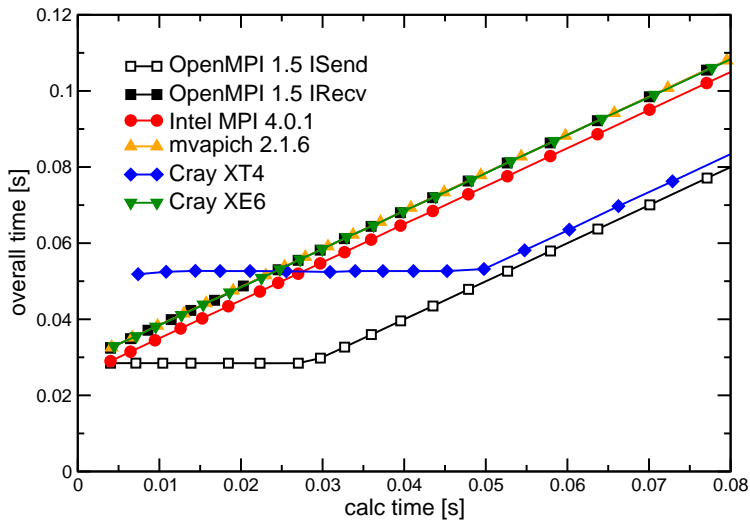}
\caption{\label{fig:ol-coca}Internode results for the nonblocking MPI benchmark
  on the Westmere-based test cluster and on Cray XT4 and XE6 systems. 
  Unless indicated otherwise, results for nonblocking
  send and receive are almost identical.\bigskip}
\end{SCfigure}

Very simple benchmark tests can be used to find out whether 
the nonblocking point-to-point communication calls in an MPI 
library do actually support truly asynchronous transfers.
Listing~\ref{lst:ol} (adapted from~\cite{hpc4se}) 
shows an example where an \verb.MPI_Irecv().
operation is set off before a function (\verb.do_work().) performs 
register-only operations for a configurable amount of time. If 
the nonblocking message transfer
overlaps with computations, the overall runtime of the code will
be constant as long as the time for computation is smaller than the
time for message transfer. We have used a constant large message length of 
80\,\MB\ to get accurate measurements. Note that the results of such tests 
may depend crucially on tunable
parameters like, e.g., the message size for the cross-over from
an ``eager'' to a ``rendezvous'' protocol, especially for small
messages. For the application
scenarios described later, 
most messages are still beyond such limits.
\begin{lstlisting}[caption={A simple benchmark to determine the capability of the MPI library to perform asynchronous nonblocking point-to-point communication for large messages (receive variant).},label=lst:ol,float=tbp]
  if(rank==0) { 
    stime = MPI_Wtime();
    %MPI\verb._.Irecv(rbuf,mcount,MPI\verb._.DOUBLE,1,0,MPI\verb._.COMM\verb._.WORLD,\&req);%
    %do\verb._.work(calctime);% 
    %MPI\verb._.Wait(req, \&status);% 
    etime = MPI_Wtime(); 
    cout << calctime << " " << etime-stime << endl; 
  } else %MPI\verb._.Send(sbuf,mcount,MPI\verb._.DOUBLE,0,0,MPI\verb._.COMM\verb._.WORLD);% 
\end{lstlisting}
Figure~\ref{fig:ol-coca} shows overall runtime versus
time for computation on the Intel Westmere cluster and the Cray XT4/XE6 systems,
respectively. We only report internode results, since no current MPI
implementation on any system supports asynchronous nonblocking intranode
communication.

On the Intel cluster we compared three different MPI implementations:
Intel MPI, OpenMPI, and MVAPICH2. The latter was compiled with the 
\verb.--enable-async-progress. flag. OpenMPI 1.5 supports a similar
setting, but it is documented to be still under development in the
current version (1.5.3), and
we were not able to produce a stable configuration with progress
threads activated. The results show that only OpenMPI (even without
progress threads explicitly enabled) was capable of
asynchronous nonblocking communication, albeit only when sending data
via a nonblocking send. The nonblocking receive is not
asynchronous, however.

Comparing the Cray XT4 and XE6 systems, it is striking that only the older
XT4 has an  MPI implementation that supports asynchronous
nonblocking transfers for large messages. 

In summary, one must conclude that the naive assumption that ``nonblocking''
and ``asynchronous'' are the same thing cannot be upheld for most current
MPI implementations; as a consequence, overlapping computation with
communication is often a matter of explicit programming. 

\subsection{MPI-parallel sparse MVM}

In the following
sections we will contrast the ``naive'' overlap via nonblocking MPI
with an approach that uses a dedicated OpenMP thread for 
explicitly asynchronous transfers. We adopt the nomenclature 
from~\cite{rw03} and \cite{hpc4se} and distinguish between ``vector mode''
and ``task mode.''

MPI parallelization of spMVM is generally done by distributing the
nonzeros (or, alternatively, the matrix rows), the right hand side
vector \verb.B(:)., and the result vector \verb.C(:). evenly across
MPI processes. Due to off-diagonal nonzeros, every process requires
some parts of the RHS vector from other processes to complete its own
chunk of the result, and must send parts of its own RHS chunk to
others. Note that it is generally difficult to 
establish good load balancing for computation and communication
at the same time. Unless indicated otherwise we use a balanced distribution of nonzeros
across the MPI processes here. At a given number of processes, 
the resulting communication pattern depends only on the
sparsity structure, so the necessary bookkeeping needs to be done only
once. 
The actual spMVM computations can be performed either by a single thread 
or, if threading is available, by
multiple threads inside the MPI process.

\begin{figure}[tb]\centering
  \includegraphics[width=0.9\linewidth]{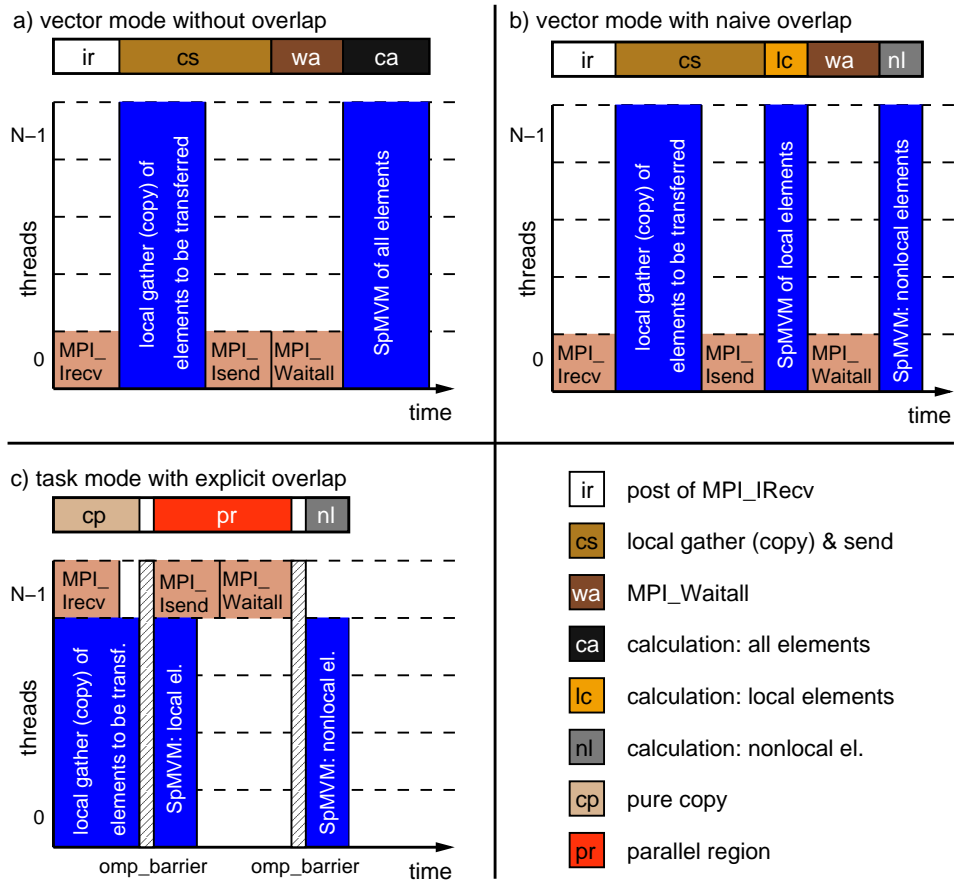}
  \caption{\label{fig:kernel_schemes}Schematic timeline view of the
    implemented hybrid kernel versions: (a) no
    communication/calculation overlap, (b) naive overlap using nonblocking
    MPI, and (c) explicit overlap by a dedicated communication thread.
    The abbreviations in the top bars indicate the individual
    contributions that will be discussed in the following sections. 
}
\end{figure}

%
%

\subsection{Vector-like parallelization: Vector mode without overlap}\label{sec:vectormode}

Gathering the data to be sent by a process 
into a contiguous send buffer may be done after the receive has
been initiated, potentially hiding the cost of copying (see
Fig.~\ref{fig:kernel_schemes}\,(a)). We call this naive approach
``hybrid vector mode,'' since it strongly resembles the
programming model for vector-parallel computers~\cite{rw03}: The time-consuming
(although probably parallel) computation step does not overlap with
communication overhead.  This is actually how ``MPI+OpenMP hybrid
programming'' is still defined in most publications. The question whether
and why using multiple threads per MPI process may improve performance
compared to a pure MPI version on the same hardware is not easy to
answer. Depending on the problem, different aspects
come into play, and there is no general rule~\cite{pdp09}.

\subsection{Vector-like parallelization: Vector mode with naive overlap}
\label{sec:vector_ol}

As an alternative one may consider hybrid vector mode with nonblocking
MPI (see Fig.~\ref{fig:kernel_schemes}\,(b)) to potentially overlap
communication with the part of spMVM that can be completed using local
RHS elements only.  After the nonlocal elements have been received,
the remaining spMVM operations can be performed. A disadvantage of
splitting the spMVM in two parts is that the local result vector must
be written twice, incurring additional memory traffic. The performance
model (\ref{eq:pmodel}) can be modified to account for an additional data transfer of
$16/N_\mathrm{nzr}$ bytes per inner loop iteration, leading to
a modified code balance of 
\bq\label{eq:mpmodel}
B^\mathrm{split}_\mathrm{CRS}
=\left(6+\frac{20}{N_\mathrm{nzr}}+\frac{\kappa}{2}\right) \frac{\bytes}{\flop}\eos
\eq
For $N_\mathrm{nzr}\approx 7\ldots 15$ and assuming $\kappa=0$,
one may expect a node-level performance penalty between
15\% and 8\%, and even less if $\kappa>0$.

For simplicity we will also use the term ``vector mode'' for pure
MPI versions with single-threaded computation.

\subsection{Task mode with explicit overlap}\label{sec:taskmode}

A safe way to ensure overlap of communication with computation is to
use a separate communication thread and leave the computational loops
to the remaining threads. We call this ``hybrid task mode,'' because
it establishes a functional decomposition of tasks (communication vs.
computation) across the resources (see
Fig.~\ref{fig:kernel_schemes}\,(c)): One thread executes MPI calls
only, while all others are used to copy data into send buffers,
perform the spMVM with the local RHS elements, and finally (after all
communication has finished) do the remaining spMVM parts.  Since spMVM
saturates at about 3--5 threads per locality domain (as shown in
Fig.~\ref{fig:nodeperf}\,(b)), at least one core per LD is available
for communication without adversely affecting node-level performance.
On architectures with SMT, like the Intel Westmere, there is also the
option of using one compute thread per physical core and bind the
communication thread to a logical core. Note that, even with perfect
overlap, one may expect the speedup compared to any vector mode
to be always less than a factor of two. 

Apart from the additional memory traffic due to writing the result
vector twice (see Sect.~\ref{sec:vector_ol}), another drawback of
hybrid task mode is that the standard OpenMP loop worksharing
directive cannot be used, since there is no concept of ``subteams'' in
the current OpenMP standard. Work distribution is thus implemented explicitly,
using one contiguous chunk of nonzeros per compute thread.

\section{Internode performance results and discussion}\label{sec:performance}

In this section we 
present strong scaling results for the Holstein-Hubbard (both basis numberings)
and sAMG matrices. Besides a discussion of the benefits of hybrid task mode we also
provide evidence that hybrid vector mode, even without overlap, may improve
performance due to better load balancing.


\subsection{Basis ordering for the Holstein-Hubbard matrix}

\begin{figure}[tb]
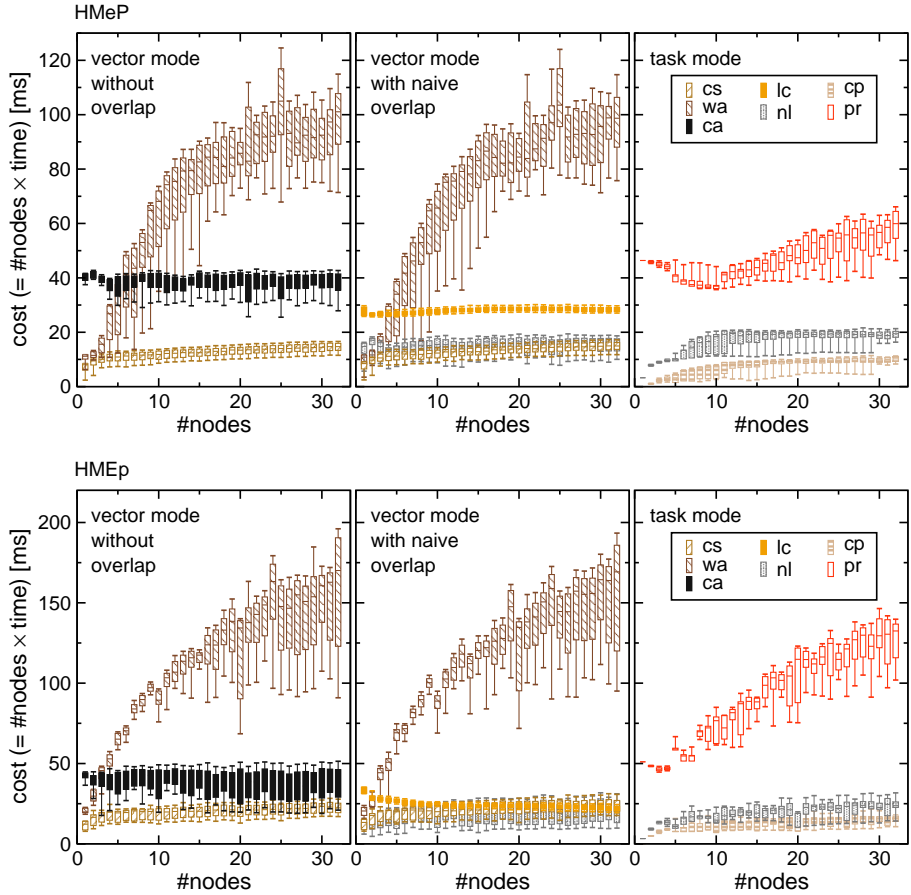
\centering
\includegraphics[width=0.85\linewidth,clip]{Contributions_rrze3vv_NZE_PPL}\\[3mm]
\includegraphics[width=0.85\linewidth,clip]{Contributions_rrze3_ROW_PPL}
\caption{\label{fig:scaling_contributions}
  Cost of the contributions to parallel spMVM vs.
  number of nodes for HMeP (top row, constant number of nonzeros per process) 
  and HMEp (bottom row, constant number of rows per process) on the
  Westmere cluster. 
  All physical cores of a node were used, either by
  MPI processes (vector modes) or additional OpenMP threads (task mode).
  Results are given in terms of a 'cost' function, which is
  the product of time required and the number of nodes.
  See Fig.~\ref{fig:kernel_schemes} for a description of the
  abbreviations.
  Costs for posting the $\texttt{MPI\_IRecv}$ are marginal
  on this scale.
}
\end{figure}
Despite the different sparsity patterns of HMeP and HMEp 
their node-level performance differs only by roughly 10\% 
(HMEp: 3.89\,\GFS, HMeP: 4.34\,\GFS\ on a Westmere EP node).
The question arises whether
it is possible to choose an appropriate partitioning of the matrix
(or, equivalently, a certain number of processes) so that
communication overhead is greatly reduced, and whether the
basis ordering plays a relevant role.
An analysis of the individual contributions of the parallel spMVM
shows the paramount role assumed by the sparsity pattern as soon as
communication becomes an issue.
In Fig.~\ref{fig:scaling_contributions}
we show for each number of nodes and one MPI process per core
(vector modes) or one MPI process per node (task mode) 
the cost for computing and communicating (time $\times$ number of nodes); each 
box with whiskers
denotes the variation across all processes in the parallel run
(10th/90th and 25th/75th percentiles). The broadening of the boxes
and whiskers
with increasing node count is a consequence of load imbalance; 
see Sect.~\ref{sec:hhmlb} for details about this issue. 

\subsubsection{Vector modes}

The purely computational cost (ca, lc, nl) is roughly on par in both matrices,
and scales almost linearly with the number of nodes (approximately
horizontal trend for the slowest processes in the cost plot), no matter 
which variant of vector mode is chosen. 
In contrast, the cost for communication spent in 
$\texttt{MPI\_Waitall}$ grows with the number of processors, which 
is to be expected since the local matrix blocks 
become smaller.
%
%
Furthermore, the inhomogeneous matrix structures result in increasing
variations of the transferred data volume (and thus communication
time).
While both matrices show this trend, there are particular
node counts for which the communication pattern of HMEp
is obviously more favorable (10, 15, 20).
At these points the number of cores is commensurable with the 
diagonal block structure of the matrix, the majority of the 
communication happens inside the nodes, and only few large messages
are passed between nodes.  
But even then the communication cost for HMEp
is still larger than for HMeP.

\subsubsection{Task mode}

In task mode, which was used here with one MPI process per node,
the overall cost is dominated by the parallel region (pr).
The reduced number of communicating MPI processes 
alleviates the load balancing problem in the communication scheme, 
which will be discussed in the following section.
More importantly, up to around 15 nodes the cost for the parallel region 
is roughly constant in the HMeP case,
implying that  communication is hidden completely behind
computation as discussed in Sect.~\ref{sec:taskmode}.
Beyond this point, communication time starts to become dominant at 
least for some processes, as indicated by the slow rise of the top
whisker (90th percentile) for the parallel region. However, we still
expect decent performance scaling for this setup.
The HMEp matrix, due to its unfavorable communication pattern,
does not allow for sufficient overlap.



\subsection{Testcase HMeP}

\subsubsection{Analysis of runtime contributions in the MPI-parallel case}\label{sec:hhmlb}

\begin{figure}[tbp]\centering
\includegraphics[width=0.85\linewidth,clip]{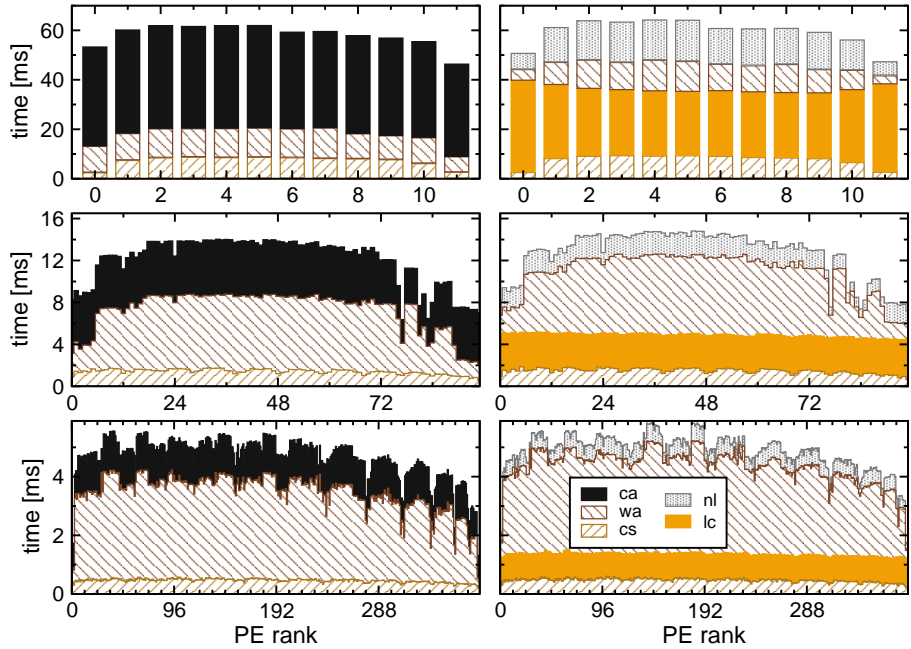}
\caption{\label{fig:balancing_bar} Contributions to the runtime of one
  spMVM (HMeP) for each MPI process at different numbers of processes
  (12, 96, and 384 from top to bottom; strong scaling)
  on the Westmere cluster. The first
  column corresponds to vector mode without overlap (see
  Fig.~\ref{fig:kernel_schemes}\,(a)) whereas the second column
  presents the runtimes for vector mode with naive overlap (see
  Fig.~\ref{fig:kernel_schemes}\,(b)). One core per process was used
  throughout. Abbreviations as in Fig.~\ref{fig:kernel_schemes}.
}
\end{figure}
In order to pinpoint the relevant performance-limiting aspects in
the MPI-parallel case we show
in Fig.~\ref{fig:balancing_bar}  the different contributions to
the runtime of a single spMVM for each MPI process when using
one of the two vector mode variants, with one process per core
and at 12, 96, and 384 processes, respectively. In these
graphs the overall runtime is always given by the highest bar;
variations in runtime across processes are a sign of load imbalance.
Owing to the separation of local from nonlocal spMVM parts, 
vector mode with naive overlap (right column) is always slower
than vector mode without overlap (left column).

The histograms reflect roughly the shape of the sparsity pattern for
HMeP as shown in Fig.~\ref{fig:mat}: As the number of processes is
increased, ``speeders,'' i.e., processes that are faster than the
rest, start to show up mainly at the top and bottom ends of the matrix
(low and high ranks). This imbalance is chiefly caused by a smaller
amount of communication (\verb.MPI_Waitall.), whereas the contribution
of computation to the runtime is much more balanced across all processes. 
One may thus expect that a lower number of MPI processes on the same
number of cores (i.e., using multiple threads per process) improves
performance due to better load balancing even without explicit overlap.
At larger process counts, execution time starts to be dominated by
communication.  Hence, even with multiple threads per MPI process
and therefore improved load balancing, explicit overlap of communication
with the local part of the spMVM (labeled ``lc'' in the right group
of diagrams in Fig.~\ref{fig:balancing_bar}) is expected to 
show significant speedups.

\subsubsection{Performance results}

\begin{figure}[tb]\centering
\includegraphics[width=0.85\linewidth,clip]{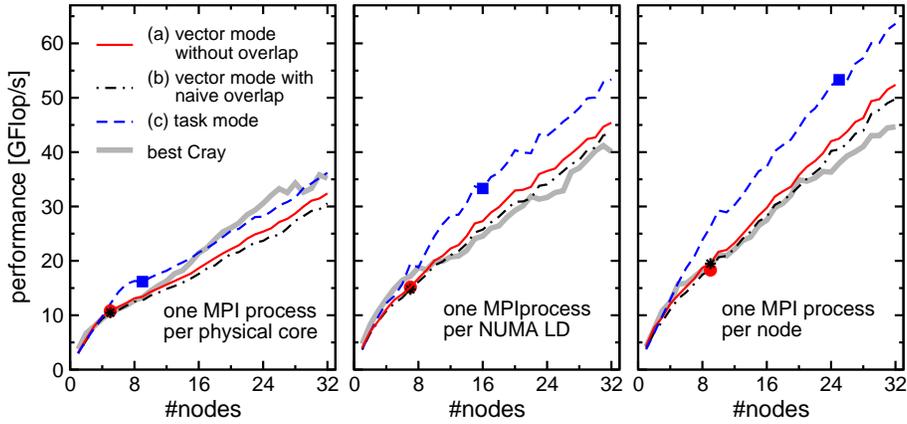}
\caption{\label{fig:nodescaling_rrze3vv}Strong scaling performance data for
  spMVM with the HMeP matrix (constant number of nonzeros per process)
  on the Intel Westmere cluster for
  different pure MPI and hybrid variants (kernel version (a) -- (c))
  as in Fig.~\ref{fig:kernel_schemes}). Symbols on each data set 
  indicate the 50\% parallel efficiency point with respect to the best
  single-node version. The best variant on the Cray 
  XE6 system is shown for reference (see text for details).}
\end{figure}
At one MPI process per physical core (left panel in 
Fig.~\ref{fig:nodescaling_rrze3vv}), vector mode with
naive overlap is always slower than the variant
without overlap because the additional data
transfer on the result vector cannot be compensated by overlapping
communication with computation.  Task mode was
implemented here with one communication thread per MPI process,
running on the second virtual core. In this case, point-to-point
transfers explicitly overlap with the local spMVM, leading to a
noticeable performance boost.  One may conclude that MPI libraries
with support for progress threads could follow the same strategy and
bind those threads to unused logical cores, thereby
allowing overlap even with
single-threaded user code.

With one MPI process per NUMA locality domain (middle panel)
the advantage of task mode is even more pronounced. Also 
the plain vector mode without overlap shows some notable speedup
compared to the MPI-only version, which was expected from
the discussion of load balancing in the previous section.
Since the memory bus of an LD is already saturated with four threads,
it does not make a difference whether six worker threads are
used with one communication thread on a virtual core, or whether
a complete physical core is devoted to communication. The same is
true with only one MPI process (12 threads) per node (right
panel).

The symbols in Fig.~\ref{fig:nodescaling_rrze3vv} indicate the 50\%
parallel efficiency point (with respect to the best single-node
performance as reported in Fig.~\ref{fig:nodeperf}\,(b)) on each data
set. In practice one would not go beyond this number of nodes because
of bad resource utilization.  For the matrix and the system under
investigation it is clear that task mode allows strong scaling to much
higher levels of parallelism with acceptable parallel efficiency than
any variant of vector mode. However, even the vector mode variants
show a significant performance advantage at multiple threads per process
due to improved load balancing.

Contrary to expectations based on the single-node performance numbers
(Fig.~\ref{fig:nodeperf}\,(b)), the Cray XE6 can generally not match
the performance of the Westmere cluster at larger node counts, with
the exception of pure MPI where both are roughly on par (left panel,
Cray results for vector mode with naive overlap). When using threaded
MPI processes (middle and right panel), task
mode performs best on the Cray system. The advantage over the other
kernel variants is by far not as pronounced as on Westmere, however.  
We have
observed a strong influence of job topology and machine load on the
communication performance over the 2D torus network. Since spMVM
requires significant non-nearest-neighbor communication with growing
process counts, the nonblocking fat tree network on the Westmere
cluster seems to be better suited for this kind of problem.  The
presented results are best values obtained on a dedicated XE6 machine.

There is also a universal drop in scalability beyond about six
nodes, which is largely independent of the particular hybrid
mode. This can be ascribed to a strong decrease in overall
internode communication volume when the number of nodes is small.
The effect is somewhat less pronounced for pure MPI, since the overhead
of intranode message passing cannot be neglected.

\subsection{Testcase sAMG}

\subsubsection{Analysis of runtime contributions in the MPI-parallel case}

\begin{figure}[tbp]\centering
\includegraphics[width=0.85\linewidth,clip]{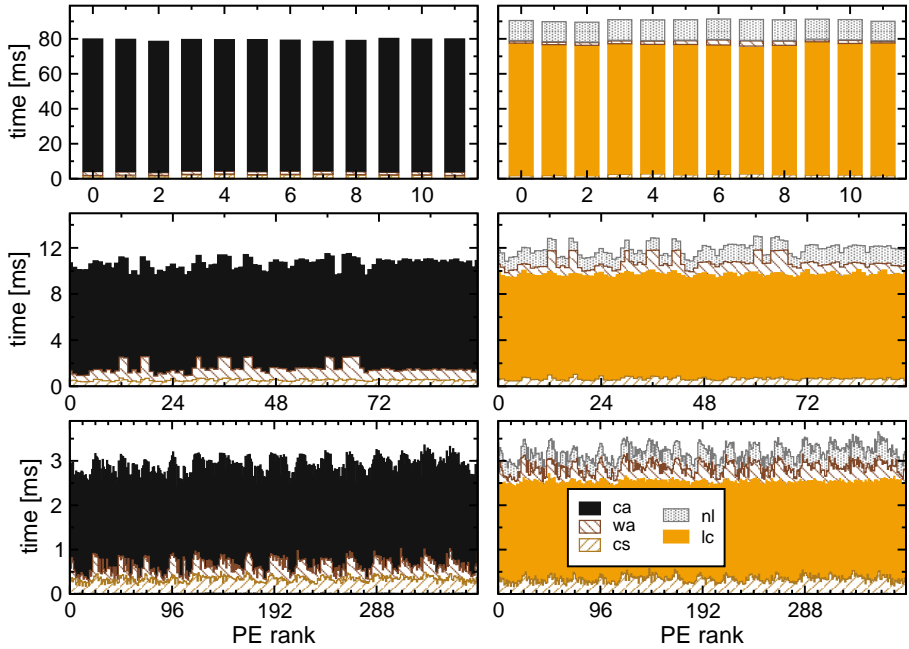}
\caption{\label{fig:balancing_bar_scai2}
  Contributions to the runtime of one spMVM (sAMG) for each MPI process.
  Parameters and abbreviations as in Fig.~\ref{fig:balancing_bar}.}
\end{figure}
As shown in Fig.~\ref{fig:balancing_bar_scai2}, there is only a slight load
imbalance for the sAMG matrix even at 384 processes
(lower diagrams). We thus only expect a marginal performance benefit
from using threaded MPI processes. Also the overall fraction of
communication overhead is quite small; task mode with explicit
overlap of communication will hence not lead to a significant
speedup.

\subsubsection{Performance results}

\begin{figure}[tb]\centering
\includegraphics[width=0.85\linewidth,clip]{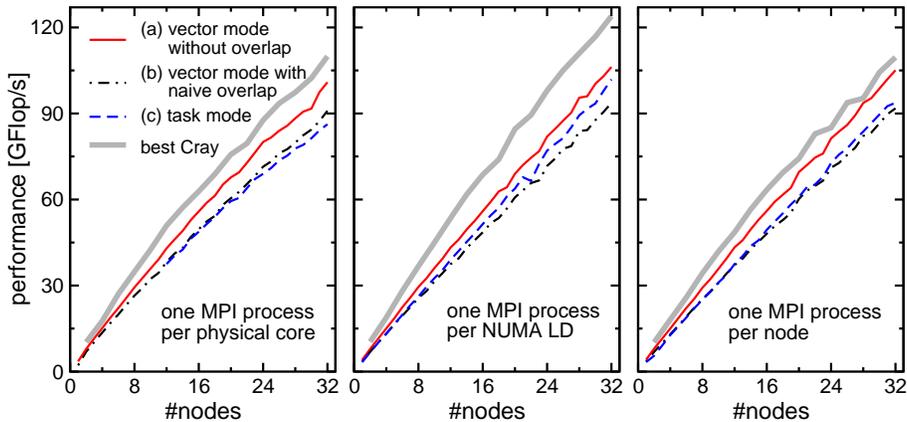}
\caption{\label{fig:nodescaling_scai2}Strong scaling performance data for
  spMVM with the sAMG matrix (same variants as in 
  Fig.~\ref{fig:nodescaling_rrze3vv}). Parallel efficiency is above 50\%
  for all versions up to 32 nodes. The Cray system performed best 
  in vector mode without overlap for all cases.}
\end{figure}
As expected from the analysis in the previous section, all variants
and hybrid modes (pure MPI, one process per LD, and one process per
node) show similar scaling behavior on the Westmere cluster,
and there is no advantage of task
mode over vector mode without overlap or over pure MPI (see
Fig.~\ref{fig:nodescaling_scai2}).  This observation supports the
general rule that it makes no sense to consider MPI+OpenMP hybrid
programming if the pure MPI code already scales well and behaves in
accordance with a single-node performance model.

On the Cray XE6, vector mode without overlap performs best across
all hybrid modes, with a significant advantage when running one
MPI process with six threads per LD. This aspect is still to be 
investigated.

\section{Summary and outlook}

We have investigated the performance properties of  pure MPI
and hybrid MPI+OpenMP hybrid variants of sparse matrix-vector multiplication
on two  multicore-based parallel systems, using  matrices
with  different sparsity patterns.  The single-node
performance  analysis on Intel Westmere and AMD Magny Cours
processors showed that memory-bound sparse MVM saturates the memory
bus of a locality domain already at about four threads, leaving
free cores for  explicit computation/communication
overlap. As most current  MPI libraries do not
support truly asynchronous point-to-point transfers, explicit
overlap enabled substantial performance gains for strong scaling
of communication-bound problems. Since the communication thread
can run on a virtual core, MPI implementations could use the same
strategy for internal ``progress threads'' and so enable asynchronous
progress without changes in MPI-only user code.

We have also identified the clear advantage of using threaded MPI
processes, even without explicit communication overlap, in cases where
computation is well balanced and load imbalance is caused by the
communication pattern.

\section*{Acknowledgments}
We thank J.~Treibig, R.~Keller and T.~Sch{\"o}nemeyer for valuable
discussions, A.~Basermann for providing and supporting the RCM
transformation and the UHBR test case as well as K.~St{\"u}ben and
H.\,J.~Plum for providing and supporting the sAMG test
case. We acknowledge financial support from KONWIHR II (project
HQS@HPC II) and thank CSCS Manno for granting access to their Cray
X6E system\@.

\end{document}